\documentstyle[12pt]{article}

\newcommand{\be}{\begin{equation}}
\newcommand{\ee}{\end{equation}}
\newcommand{\ba}{\begin{eqnarray}}
\newcommand{\ea}{\end{eqnarray}}

\begin{document}
\begin{flushright}
Hep-th/yymmxxx
\end{flushright}  
\begin{center}
\Large{REISSNER--NORDSTR\"OM TYPE BLACK HOLES 
IN DILATON--AXION GRAVITY}
\end{center}
\vskip1cm
\begin{center}
{\bf D. V. Gal'tsov\footnote{Permanent 
address: Department of Theoretical Physics, Physics Faculty, 
Moscow State University, 119899, Moscow, Russia; 
e-mail: galtsov@grg.phys.msu.su}
and  P. S. Letelier\footnote{e-mail: letelier@ime.unicamp.br}}
\\
Departamento de Matem\'{a}tica Aplicada -- IMECC\\
Universidade Estadual de Campinas\\                   
13081 Campinas, S.P., Brazil\\
\end{center}

\centerline{\bf Abstract}

\begin{quotation}
\vskip .5 cm

A $2p + 5$ parametric family of  black holes  is 
constructed in dilaton--axion gravity with $p$ vector fields using
a holomorphic representation of U--duality in three dimensions.   
The metric of the non--extremal black holes has a Reissner--Nordstr\"om 
type structure and generically possesses an internal horizon. However 
in the extremal limit the generic solution exhibits a dilatonic
type null singularity. Only in the case of the orthogonal electric
and magnetic charges (if $p>1$) the extremal solution may have a
 non--singular event horizon. 
\end{quotation}

%\newpage

Black holes in theories including dilaton and axion fields (N=4 supergravity,
low--energy heterotic string effective action) has attracted recently
much attention. Static black holes in N=4 supergravity were discussed some
 time ago by Gibbons \cite{gi} and Gibbons and Maeda \cite{gim} who
 found that the dilaton
changes substantially the nature of a charged black hole, transforming it
from the Reissner--Nordstr\"om to ``dilatonic'' type with no internal horizon.
This solution was rediscovered in \cite{ghs} in a particularly
simple gauge making this property explicit. It was realized later
that if the axion is switched off, the {\it dyon} solutions may still
have a Reissner--Nordstr\"om structure \cite{rk4}. To be consistent with
N=4 supergravity, one has to assume that electric and magnetic charges
belong to different $U(1)$ sectors, the axion being non excited. 
When the axion {\it is} excited, the situation becomes more complicated.
Recently it was found that in the dilaton--axion gravity with multiple 
vector fields there exist extremal solutions of the Israel--Wilson--Perjes 
type \cite{bko} similar to well-known solutions of the Einstein--Maxwell 
theory. However, general spherically--symmetric black hole 
solutions to dilaton--axion gravity with multiple vector fields seem not 
to be given so far apart from the BPS limit.

Here, using the solution--generating technique based on the Ehlers--Harrison
form of symmetry transformations \cite{gk}, we find a very simple form
of the general static solution which has the metric of the Reissner--Nordstr\"om
type with ``variable mass''. It interpolates between the usual Reissner--Nordstr\"om
and dilatonic solutions to which it reduces in different particular cases.
 
The action we are dealing with is
\begin{equation}  
S=-\frac{1}{16\pi}\int \left\{R+\frac{2 \nabla z \nabla{\bar z}}
{(z-{\bar z})^{2}} -
2{\rm Re}\left(iz{\cal F}_{\mu\nu}^a{\cal F}^{a\, \mu\nu}\right)\right\}
\sqrt{-g}d^4x,
\end{equation}
where $z=\kappa+i{\rm e}^{2\phi}$ is the complex axidilaton field, 
${\cal F}^a_{\mu\nu}=(F^a_{\mu\nu}+i{\tilde F}^a_{\mu\nu})/2,\; 
{\tilde F}^{a\, \mu\nu}=\frac{1}{2}E^{\mu\nu\lambda\tau}
F^a_{\lambda\tau},\; a=1,...,p$ describe the
set of $p\;U(1)$ vector fields. We use the standard ansatz for the line
element of the stationary spacetime metric   
\begin{equation}
ds^2=g_{\mu\nu}dx^\mu dx^\nu=f(dt-\omega_idx^i)^2-\frac{1}{f}h_{ij}dx^idx^j,
\end{equation}
where the three--space metric $h_{ij}$, the covector  
$\omega_i,\; (i, j=1, 2, 3)$, and the scalar $f$ depend on the space 
coordinates $x^i$ only. The vector fields give rise to the sets of 
electric $v^a$ and magnetic $u^a$ potentials via
\be
F^a_{i0}=\frac{1}{\sqrt{2}}\partial_i v^a,\quad 
{\tilde F}^a_{i0}=\frac{{\rm e}^{2\phi}}{\sqrt{2}}
(\kappa\partial_iv^a-\partial_iu^a).
\ee
The  resulting three--dimensional theory is equivalent to the 
gravity--coupled $\sigma$--model possessing a K\"ahler target manifold
$SO(2, 2+p)/SO(2)\times SO(1,1+p)$. Holomorphic coordinates include
the axidilaton $z$, the set of complex electromagnetic potentials
$\Phi^a$ and the generalized Ernst potential $E$ \cite{ad}:
\be
\Phi^a=u^a-zv^a,\quad E=if-\chi+v^a\Phi^a,
\ee
where $\chi$ is the twist (NUT) potential. Here and in what follows 
a sum from $1\;{\rm to}\;p$ is assumed over repeated $SO(p)$ 
indices. The isometry group of the
target manifold can be used to generate the solutions to the
theory (1) starting with solutions to the vacuum Einstein equations.
Here we derive a fully $SO(2, 2+p)$ covariant counterpart of the
Schwarzschild solution, describing generic spherically
symmetric black holes in the dilaton--axion gravity with multiple
vector fields.
 
The total set of $(p^2+7p+12)/2 \; SO(2, 2+p)$ transformations can
be decomposed  as follows. There are $2p$ gauge transformations 
fixing $v^a(\infty)=0,\,u^a(\infty)=0$; two transformations ensuring 
an asymptotic flatness $f(\infty)=1,\,\chi(\infty)=0;\; (p^2-p)/2$  
elements of the residual $SO(p)$ invariance; and the
rest of $2p+4$ transformations which may bring new parameters 
to  the solution (as we will see one of them is still a gauge parameter
 which is worth to be kept free). 
Together with the mass $M_0$ of the seed Schwarzschild
metric this gives the maximal number of $2p+5$ free parameters. 
These can be conveniently chosen as two charge $SO(p)$ vectors
$\mu^a,\,\nu^a$, the mass $M$, two parameters $b,c$ defining one of 
the scalar charges (or any their combination) and the NUT charge,
and the asymptotic values of the dilaton $\phi_{\infty}$
and axion $\kappa_{\infty}$.

Magnetic charges  $\mu^a$ are introduced via a (magnetic)
Harrison transformation subject to asymptotic conditions  
$z(\infty)=E(\infty)=i,\; \Phi^a(\infty)=0$:
\be
E\rightarrow E\Pi^{-1},\quad
z\rightarrow z\Pi^{-1},\quad
\Phi^a\rightarrow \left\{\Phi^a+\mu^a(\Lambda+1)\right\}\Pi^{-1},
\ee
where $\Pi=1+2\mu^a\Phi^a+\mu^2\Lambda,\; \Lambda=zE+(\Phi^a)^2,\;
\mu^2=(\mu^a)^2$. Similar electric charging transformation 
reads:
\[
E\rightarrow (1-\nu^2)^{-1}\left(E+2\nu^a\Phi^a-\nu^2 z\right),\quad
z\rightarrow (1-\nu^2)^{-1}\left(z-2\nu^a\Phi^a-\nu^2 E\right),\]
\be
\Phi^a\rightarrow \Phi^a+\nu^a(1-\nu^2)^{-1}
\left(E+2\nu^b\Phi^b -z\right), 
\ee
where $\nu^2=(\nu^a)^2$. An arbitrary NUT charge is 
generated by the Ehlers transformation, 
while one of the scalar charges can be produced using an analogous 
element of the $S$--duality subgroup. The combined transformation 
including two real parameters $c,\, b$ (the first responsible for 
NUT, the second for a scalar charge) reads
\ba
&E\rightarrow \left\{(E-c)\Pi_1+b(1+c^2)(\Phi^a)^2\right\}
(1+cE)^{-1}\Pi_1^{-1},&\nonumber\\
&z\rightarrow \left(z+c\Lambda-b-bcE\right)\Pi_1^{-1},\quad
\Phi^a\rightarrow \Phi^a (1+c^2)^{1/2}(1+b^2)^{1/2}\Pi_1^{-1}&,
\ea
where $\Pi_1=1+c\Lambda+bz+bc\Lambda$. Finally, free asymptotic values 
of the dilaton and axion are generated by
\be
z\rightarrow e^{-2\phi_\infty}z + \kappa_\infty ,\quad 
\Phi^a\rightarrow e^{-\phi_\infty}\Phi^a,
\ee
with $E$ unchanged.

In order to covariantize the Schwarzschild solution with respect to
the maximal  subgroup of the three--dimensional $U$--duality group, 
preserving an asymptotic flatness, 
we apply all these transformations (in the above order) to the seed
Schwarzschild metric $f_0=1-2M_0/r_0,\; h_{rr}=1,\, h_{\theta\theta}=
h_{\varphi\varphi}\sin^{-2}\theta=f_0 r_0^2$.   
The resulting solution in terms of complex potentials reads:
\ba
& E=\left\{(1-f_0)\left[2q(1+bc)+n(b-c)\right]+im\left({\tilde n}_f
-bcn_f\right)-mn(b+cf_0)\right\}G^{-1},&\nonumber\\
& z=\kappa_{\infty}-e^{-2\phi_\infty}\left\{2q(1-f_0)(1+bc)+
n(c{\tilde m}_f+bm_f)-
im(n_f-bc{\tilde n}_f)\right\}G^{-1},&\nonumber\\
&\Phi^a=e^{-\phi_\infty}(1+c^2)^{1/2}(1+b^2)^{1/2}(1-f_0)
\left(\mu^a n+\nu^a(2q-im)\right)G^{-1},&
\ea
where
\ba
&q=\mu^a\nu^a,\;\;n=1-\nu^2,\;\; m=1-\mu^2, &\nonumber\\
&m_f=1-\mu^2f_0, \;\;n_f=1-\nu^2f_0,\;\;{\tilde m}_f=f_0-\mu^2,\;\;
{\tilde n}_f=f_0-\nu^2,&\nonumber\\
&G=2q(1-f_0)(c-b)+im(bn_f+c{\tilde n}_f)+n(m_f-bc{\tilde m}_f).&
\ea
This solution satisfies the asymptotic flatness condition $E(\infty)=i$
and is normalized so that $\Phi^a(\infty)=0,\; \phi(\infty)=\phi_\infty,\;
\kappa(\infty)=\kappa_\infty$. It is suitable, in particular, to probe recent 
suggestions by Gibbons, Kallosh and Kol \cite{gkk}. We do not, however, intend
to discuss this in the present paper, so, for simplicity, we set 
$\phi_\infty=\kappa_\infty=0$ in what follows.  

Now let us define complex charges ${\cal M}=M+iN$, (mass and NUT), 
${\cal D}=D+iA$, (dilaton and axion), ${\cal Q}^a=Q^a+iP^a$ 
(electric and magnetic), via the asymptotics  
\be
E\sim i\left(1-\frac{2{\cal M}}{r}\right),\quad
z\sim i\left(1-\frac{2{\cal D}}{r}\right),\quad
\Phi^a \sim -i\frac{\sqrt{2}{\cal Q}^a}{r},
\ee
where $r$ is some radial variable (coinciding with $r_0$ asymptotically).
>From (9) one obtains
\[
{\cal M}=\frac{M_0}{mn} e^{2i\varphi_c}\left(\mu^2\nu^2-2iq-1\right),\quad
{\cal D}=\frac{M_0}{mn} e^{2i\varphi_b}\left(\nu^a+i\mu^a\right)^2,\]
\be
{\cal Q}^a=-\frac{\sqrt{2}M_0}{mn} e^{i(\varphi_b+\varphi_c)}
\left(m\nu^a+i(n\mu^a+2q\nu^a)\right),
\ee
where $\varphi_c=\arctan(1/c),\, \varphi_b=\arctan(1/b)$. Hence $c$ rotates
mass and NUT parameter, while
$b$ rotates dilaton and axion charges. Both rotations are accompanied
by suitable transformations of electric and magnetic charges. Independently
on the choice of $c$ and $b$ the following holomorphic relation between
complex charges holds
\be
{\cal D}=-\frac{({\cal Q}^a)^2}{2{\cal M}}.
\ee
This is a covariant (with respect to the full $U$--duality group) 
generalization to many--vector case 
of the well-known relation between scalar and vector charges in
dilaton--axion gravity \cite{gk}. The new feature now is that, by a suitable 
choice of $b$, one can always arrange, say, for a zero value of the axion (dilaton)
charge. This does not contradict to the fact that scalar charges are
determined by the values of vector charges since the latter will also
be affected by the choice of $b$. Similarly, by an appropriate choice of $c$
one can annihilate the NUT charge $N$ or mass $M$ of the solution. 
Note that $c$--transformation does not affect the scalar charges, but it 
does change vector ones to comply with (13).

In view of (13), the total number of independent Coulomb charges 
defining our solution is $2p+2$, namely, $M, N, Q^a, P^a$. Since
the solution can be regarded as a harmonic map to a symmetric space,
it should be uniquely determined by the charges. Note, that no relation between
$2p+3$ parameters $M_0, \mu^a, \nu^a, c, b$ follows from (13), which is merely 
an identity in terms of the parametrization (12). Thus one of parameters  
$M_0, \mu^a, \nu^a, c, b$ is pure gauge, the natural choice being $b$. 
It is, however, convenient to keep $b$ as a free parameter
because this provides a simple way to construct solutions with a desired
value of one of scalar charges.
 
The metric function $f$ entering the line element  (2) depends 
on the parameter $c$ and {\it does not} depend on $b$:
\be
f=f_0 mn (1+c^2)F^{-1}, \quad F=m_f n_f+c^2{\tilde n}_f{\tilde m}_f+2cq
(1-f^2_0).
\ee
Also, as could be expected, $f$ depends on electric and magnetic charge 
parameters $\nu^a,\,\mu^a$ only through three $SO(p)$ invariants 
$\mu^2, \nu^2, q$. This reflects the fact that by a suitable $SO(p)$ 
rotation one can always choose, say, the vector $\nu^a$ to have only one
non--zero component, then using the residual $SO(p-1)$, preserving 
the direction of $\nu^a$,  one can annihilate $p-2$ components of $\mu^a$, 
so only three of $2p$ components  are essential (the rest $SO(p-2)$ being pure
gauge). Note, that the denominator $F$ of $f$ is a quadratic
function of $f_0$. Since $f_0$ has one zero, this means
that in the resulting solution $g_{00}$ generally will have two zeroes, thus
exhibiting the Reissner--Nordstr\"om like behavior. To write down the metric
explicitly one has to choose coordinates in the new solution. For dilaton black
holes (non--dyons) the particularly simple choice of coordinates \cite{ghs} is
$f=1-2M/r,\;g_{\theta\theta} =r(r-r_-)$ where $r_-$ is the position of 
the singularity. No such simple gauge is possible generally for
(14), when the solution is expected to have two horizons. Therefore it is more
natural now to choose the curvature (Schwarzschild) coordinates
fixed by the condition $g_{\theta\theta} =r^2$, so that the singularity
is at $r=0$. Since the three--metric $h_{ij}$ in (2) is not affected
by the transformations, one has the following condition relating 
$r_0$ in the seed solution and $r:\; r_0^2f_0=r^2f$. Using (14) one finds
\be
r_0=\rho+\sqrt{\sigma^2+r^2},\quad \rho=\frac{r_0^+ +r_0^-}{2}, 
\quad \sigma=\frac{r_0^+ -r_0^-}{2}, 
\ee
where $r_0^\pm$ are roots of the equation $F(r_0)=0$:
\be
r_0^\pm=\frac{M_0}{mn(1+c^2)}\left(2\alpha-\beta
\pm\sqrt{\beta^2-4\alpha\gamma}\right),
\ee with
\be
\alpha=c^2-2cq+\mu^2\nu^2,\;\;\beta=(\mu^2+\nu^2)(1+c^2),\;\;
\gamma=1+2cq+c^2\mu^2\nu^2.
\ee
In curvature coordinates the transformed metric look similarly
to the Reissner--Nordstr\"om solution with a coordinate--dependent mass:
\be
ds^2=\left(1-\frac{2Mg}{r}+\frac{C}{r^2}\right)(dt-2Nd\varphi)^2-
\left(1-\frac{2Mg}{r}+\frac{C}{r^2}\right)^{-1}dr^2-
r^2\left(d\theta^2+\sin^2\theta d\varphi^2\right).
\ee
Here the mass function $g(r)$ (asymptotically normalized  to unity) is
\be
g=\sqrt{1+\frac{\sigma^2}{r^2}},
\ee 
while the physical mass $M$ of the solution and the ``charge'' 
parameter $C$  read:
\be
M=\frac{M_0\left\{(1-c^2)(1-\mu^2\nu^2)+4cq \right\}}{ mn (1+c^2)} ,\quad 
C=\sigma^2+\rho(\rho-2M_0).  
\ee

This metric describes a {\it generic} spherically--symmetric
black hole in dilaton--axion gravity with multiple vector fields 
( $N=4$ supergravity if $p=6$).
Asymptotically it is Reissner--Nordstr\"om, but now the metric function $f$
has a modified short--range behavior. The behavior at the singularity $r=0$
is similar to that of the Reissner--Nordstr\"om metric: Kretschmann scalar
diverges as $r^{-8}$.

Our new solution (14) interpolates between the standard
 Reissner--Nordstr\"om--NUT solution  ($\sigma=0$), and the dilaton
 black hole solution \cite{gi, ghs}, to which it reduces when
\be
r_0^+(r_0^-+2M)=0.
\ee
In this latter case, passing to a new radial variable $r_1=r_0-r_0^-$ and
introducing a parameter $r_-=2\sigma$, one obtains
\be
ds^2=\left(1-\frac{2M}{r_1} \right)(dt-2Nd\varphi)^2-
\left(1-\frac{2M}{r_1}\right)^{-1}dr_1^2-
r_1(r_1-r_-)\left(d\theta^2+\sin^2\theta d\varphi^2\right).
\ee

Now let us discuss a proper black hole case $N=0$. This
corresponds to the following value of $c$ :
\be
c=-\frac{1-\mu^2\nu^2}{2q}+\sqrt{1+\left(\frac{1-\mu^2\nu^2}{2q}\right)^2},
\ee
provided $q\neq 0$ (and $c=0$ for $q=0$). One can use an inverse equation
to express $q$ as a function of $c$, keeping in mind that $c$ in the resulting
expressions has a fixed value (23). Then
 \[
 C=\frac{2M^2\left\{(\mu^2+\nu^2)(1+\mu^2\nu^2)(1-c^2)^2+
4(c^2-\mu^2\nu^2)(1-c^2\mu^2\nu^2)\right\}}{(1+c^2)^2(1-\mu^2\nu^2)^2},
\]
\be
\sigma^2=\frac{M^2 \left\{(1-c^2)^2(\mu^2+\nu^2)^2+
4(c^2-\mu^2\nu^2)(1-c^2\mu^2\nu^2)\right\}}{(1+c^2)^2(1-\mu^2\nu^2)^2}.
\ee
A particularly simple form parameters of the solution take when $q=0$, 
what means that either one of two $SO(p)$ vectors $\nu^a,\, \mu^a$
vanishes, or they are orthogonal. Then the solution is NUT--less if
$c=0$. In the one--vector case $q^2=\mu^2\nu^2$, so this means that the solution
is non--dyon. In this case (21) is satisfied, and we recover the metric (22). 
Another interesting case (if $p\neq 1$) is $\mu^2=\nu^2,\, q=0$ (symmetric dyon). 
Then $\sigma =0$ and the metric is {\it pure} Reissner--Nordstr\"om. 

Generically the solution (18) has two horizons, $r^\pm$, provided  
$\mu^2\nu^2<1,\;c^2<\mu^2\nu^2$:
\be
r^\pm=\frac{2M\sqrt{mn(1-c^2)}}{(1+c^2)(1-\mu^2\nu^2)}\lambda^\pm,
\quad \lambda^+=\sqrt{1-c^2\mu^2\nu^2},\;\; \lambda^-=\sqrt{\mu^2\nu^2-c^2}, 
\ee
and possesses a spacelike singularity. It is therefore  not expressible in the
``dilatonic'' form (22), and rather is similar to the usual Reissner--Nordstr\"om
metric. Somewhat surprisingly, in the extremal limit the {\it generic} solution
is ``dilatonic''. The extremality limit corresponds to one or both $\mu^2,\,\nu^2$
approaching unity. Using (12) one can show that the BPS
bound will be then saturated indeed
\be
|{\cal M}|^2+|{\cal D}|^2-|{\cal Q}|^2=0,
\ee
and it is worth noting that the BPS limit corresponds to vanishing $M_0$, that is, 
to the {\it vacuum} seed solution.

If $q\neq 0$, and one of two parameters $\mu^2,\,\nu^2$ approaches 
unity (the other still remaining arbitrary) then both $r^\pm$ tend to zero,
and we observe typical behavior of the extremal dilaton black hole
(null singularity).  However if $q= 0$, (what in the NUT--less case
implies $c=0$), there are two possibilities. If one of parameters $\mu^2,\,\nu^2$ 
tends to unity and another not, we have the same situation. But
if $\mu^2=\nu^2$ and both approach unity, then in the BPS limit
\be
r^+=r^-=M,
\ee
that is the radius of the horizon remains finite. Thus, in the extremal  
limit, the generic behavior is {\it dilatonic}, only a restricted class  
$q=0,\, \mu^2=\nu^2=1$  of black holes has a regular horizon. Note that this 
latter option is only possible when the number of vector fields $p>1$. 
BPS solutions in the dilaton--axion gravity with multiple vector fields 
were discussed recently in detail by Kallosh {\it et al.} \cite{bko}, general
BPS solutions in the $p=1$ case were given in \cite{cg}.
It was also discovered that in the theories with moduli fields more
complicated black hole metrics appear \cite{cv} which deviate  both
from the Reissner--Nordstr\"om and dilatonic black hole solutions.
 
\vskip3mm
{\large \bf Acknowledgments}\\
This work was supported in part by FAPESP and CNPq, Brazil.  D.G.  is
grateful to Departamento de Matematica Applicada, IMECC,
Universidade Estadual de Campinas for hospitality during his visit.
The work of D.G. was also supported in part by the RFBR
Grant 96--02--18899.

\end{document}